\documentstyle[12pt]{article}

\def\as{\alpha_s}

\newcommand{\mt}{m_{\rm t}}
\newcommand{\mtb}{\overline{m}_{\rm t}}

\newcommand{\mb}{m_{\rm b}}
\newcommand{\mw}{M_{\rm W}}
\newcommand{\mz}{M_{\rm Z}}
\newcommand{\gev}{\, {\rm GeV}}

\newcommand{\bea}{\begin{eqnarray}}
\newcommand{\eea}{\end{eqnarray}}
\newcommand{\bd}{\begin{displaymath}}
\newcommand{\ed}{\end{displaymath}}

\newcommand{\beq}{\begin{equation}}
\newcommand{\eeq}{\end{equation}}
\newcommand{\be}{\begin{equation}}
\newcommand{\ee}{\end{equation}}
\newcommand{\ord}{{\cal O}}

\newcommand{\f}{\frac}

\begin{document}


\thispagestyle{empty}

\vspace*{1.2truecm}
\bigskip

\centerline{\LARGE\bf  On the Scale Uncertainties}
\vspace{0.3truecm}
\centerline{\LARGE\bf  in the $B \to X_s \gamma$ Decay}
 \vskip1truecm
\centerline{\large\bf Andrzej J. Buras, Axel Kwiatkowski and Nicolas Pott}
\bigskip
\centerline{\sl  Technische Universit{\"a}t M{\"u}nchen, Physik 
Department}
\centerline{\sl D-85748 Garching, Germany}
\vspace{1.5truecm}
\centerline{\bf Abstract}
We analyze the theoretical uncertainties in $Br(B\to X_s\gamma)$
due to the choice of the high energy matching scale $\mu_W=\ord(\mw)$
and the scale $\mu_t$ at which the running top quark mass is defined:
$\mtb(\mu_t)$. To this end we have repeated the calculation of the 
initial conditions confirming the final results of Adel and Yao and 
Greub and Hurth and generalizing them to include
the dependences on $\mu_t$ and $\mu_W$ with $\mu_t\not=\mu_W$. 
In the leading order the $\mu_W$ 
and $\mu_t$ uncertainties  in $Br(B\to X_s\gamma)$ turn out
to be $\pm 13\%$ and $\pm 3\%$ respectively. We show analytically how 
these uncertainties are reduced after including next-to-leading QCD 
corrections. They amount to $\pm 1.1\%$ and $\pm 0.4\%$ respectively.
Reanalyzing the uncertainties due to the scale $\mu_b=\ord(m_b)$
we find that after the inclusion of NLO effects 
they amount to $\pm 4.3\%$ which is a factor 2/3 smaller than
claimed in the literature.  Including 
the uncertainties due to input parameters as well as
the non-perturbative $1/m_b^2$ and $1/m_c^2$ corrections
we find $Br(B{\to}X_s \gamma) = (3.60 \pm 0.33) \times 10^{-4}$
where the error is dominated by uncertainties in the input
parameters. This should be compared with $(3.28 \pm 0.33) \times 10^{-4}$
 found by Chetyrkin et al.  where the error is shared evenly between
the scale and parametric uncertainties.  
 
\vspace*{2.0cm}

\begin{center}
{\small 
Supported by the
German Bundesministerium f{\"u}r Bildung and Forschung under contract 
06 TM 874 and DFG Project Li 519/2-2.

}
\end{center}


\newpage

\noindent
{\large\bf 1.} The inclusive $B \to X_s \gamma$ decay has been subject of
considerable experimental and theoretical interest during the last
ten years. Experimentally its branching ratio is found 
by the CLEO collaboration to be \cite{CLEO2}
\begin{equation}\label{EXP}
Br(B \to X_s\gamma) = (2.32 \pm 0.57 \pm 0.35) \times 10^{-4}\,,
\label{incl}
\end{equation}
and a very recent preliminary result from the ALEPH collaboration
reads \cite{ALEPH}
\begin{equation}\label{EXP2}
Br(B \to X_s\gamma) = (3.38 \pm 0.74 \pm 0.85) \times 10^{-4}.
\end{equation}
In (\ref{EXP}) and (\ref{EXP2})
 the first error is statistical and the second is systematic.
On the other hand the complete NLO analysis
gives \cite{CZMM} 
\be\label{theon}
Br(B{\to}X_s \gamma) = (3.28 \pm 0.22~({\rm scale})~\pm 0.25~({\rm par})) 
  \times 10^{-4} =
 (3.28 \pm 0.33) \times 10^{-4}.
\ee
where the first error results from the scale uncertainty (see below) and
the second error from the uncertainties in the input parameters.
A similar result has been obtained in \cite{GREUB}.

\noindent
The NLO analyses presented in \cite{CZMM,GREUB} reduced by a 
factor of 3-4 the $\mu_b$-uncertainties \cite{AG1,BMMP:94} 
present in the leading order,
where $\mu_b=\ord(m_b)$ is the scale at which the relevant decay
matrix element is evaluated. This reduction of the $\mu_b$
uncertainty is very welcome because in the forthcoming years much more
precise measurements of $Br(B{\to}X_s \gamma)$ are expected from the
upgraded CLEO detector, as well as from the B-factories at SLAC and KEK.
This is also the reason why continuing efforts are being made to estimate
non-perturbative corrections to the $B{\to}X_s \gamma$ decay with
higher precision \cite{FLS96,LDGAMMA,VOL96,BUC97} as well. 
It appears that these latter corrections amount only to 
a few percent and constitute a rather small theoretical uncertainty.
 
\noindent
{\large\bf 2.} In this letter we have repeated the numerical
analysis of \cite{CZMM} to find that the remaining scale uncertainties are
by roughly a factor 1.5-2.0 smaller than quoted by these authors and in
\cite{GREUB}. This includes  also two additional 
theoretical
uncertainties which have not been addressed in the literature. They
are related to the choice of the high energy matching scale 
$\mu_W=\ord(\mw)$
and the scale $\mu_t=\ord(\mt)$ at which the running top quark mass is defined:
$\mtb(\mu_t)$. These two scales enter the analysis of $B{\to}X_s \gamma$
in the process of calculating the initial conditions for the 
renormalization group running of 
the Wilson
coefficients $C_7$ and $C_8$ of the operators

\begin{equation}\label{O6}
Q_{7}  =  \frac{e}{8\pi^2} m_b \bar{s}_\alpha \sigma^{\mu\nu}
          (1+\gamma_5) b_\alpha F_{\mu\nu}\qquad            
Q_{8}     =  \frac{g_s}{8\pi^2} m_b \bar{s}_\alpha \sigma^{\mu\nu}
   (1+\gamma_5)T^a_{\alpha\beta} b_\beta G^a_{\mu\nu}.
\end{equation}
Here $e$ and $g_s$ denote the electromagnetic and strong coupling
constants respectively.
These initial conditions have been calculated at NLO in \cite{Yao1} and
have been recently confirmed in \cite{GH97}. These NLO corrections
are necessary to remove the renormalization scheme dependence present
in the renormalization group evolution from $\mu_W=\ord(\mw)$ down
to $\mu_b=\ord(\mb)$. From our point of view the additional reason
for performing these rather tedious calculations is the reduction
of the uncertainties related to the choices of $\mu_W$ and $\mu_t$.
These uncertainties have not been discussed in 
\cite{CZMM,GREUB,Yao1,GH97}.

\noindent
To this end we have repeated the calculation of the initial condition
for the by far dominant Wilson coefficient $C_7$ confirming the 
final result in \cite{Yao1,GH97} and 
generalizing it to include
the dependences on $\mu_t$ and $\mu_W$ with $\mu_t\not=\mu_W$. 
In \cite{Yao1}  and \cite{GH97} 
$\mu_W=\mu_t=\mu_{Wt}$ have been
used. The technical details of our calculation which differs in certain
aspects from the previous ones will be presented elsewhere \cite{BKP2}.
Here we discuss first the issue of the $\mu_W$ and $\mu_t$ uncertainties
 and their
reduction after the inclusion of NLO corrections.
 Subsequently we discuss the $\mu_b$ uncertainties and we present our 
estimate of $Br(B \to X_s \gamma)$ in the Standard Model.

\noindent
{\large\bf 3.}
In the leading
logarithmic approximation
one has
\begin{equation}\label{main}
\frac{Br(B \to X_s \gamma)}
     {Br(B \to X_c e \bar{\nu}_e)}=
 \frac{|V_{ts}^* V_{tb}^{}|^2}{|V_{cb}|^2} 
\frac{6 \alpha}{\pi f(z)} |C^{(0){\rm eff}}_{7}(\mu_b)|^2\,,
\end{equation}
where
\begin{equation}\label{g}
f(z) = 1 - 8z + 8z^3 - z^4 - 12z^2 \ln z           
\quad\mbox{with}\quad
z =
\frac{m^2_{c,pole}}{m^2_{b,pole}}
\end{equation}
is the phase space factor in $Br(B \to X_c e \bar{\nu}_e)$ and
$\alpha=e^2/4\pi$.

\noindent
The effective renormalization scheme independent coefficient 
$C^{(0){\rm eff}}_{7}(\mu_b)$ introduced in \cite{BMMP:94} is given by
\begin{equation}
\label{C7eff}
C_{7}^{(0){\rm eff}}(\mu_b)  =  
\eta^\frac{16}{23} C_{7}^{(0)}(\mu_W) + \frac{8}{3}
   \left(\eta^\frac{14}{23} - \eta^\frac{16}{23}\right) 
C_{8}^{(0)}(\mu_W) +
    \sum_{i=1}^8 h_i \eta^{a_i}\,,
\end{equation}
where
\begin{equation}\label{ET}
\eta  =  \frac{\as(\mu_W)}{\as(\mu_b)},
\end{equation}
\begin{equation}\label{c7}
C^{(0)}_{7} (\mu_W) = \frac{3 x_t^3-2 x_t^2}{4(x_t-1)^4}\ln x_t + 
   \frac{-8 x_t^3 - 5 x_t^2 + 7 x_t}{24(x_t-1)^3},
\end{equation}

\begin{equation}\label{c8}
C^{(0)}_{8}(\mu_W) = \frac{-3 x_t^2}{4(x_t-1)^4}\ln x_t +
   \frac{-x_t^3 + 5 x_t^2 + 2 x_t}{8(x_t-1)^3}                               
   \end{equation}
with
\be\label{xt}
x_t=\f{\mtb^2(\mu_t)}{\mw^2}.
\ee
The numbers $a_i$ and  $h_i$  are
given in table \ref{tab:akh}.

\begin{table}[htb]
\begin{center}
\begin{tabular}{|r|r|r|r|r|r|r|r|r|}
\hline
$i$ & 1 & 2 & 3 & 4 & 5 & 6 & 7 & 8 \\
\hline
$a_i $&$ \frac{14}{23} $&$ \frac{16}{23} $&$ \frac{6}{23} $&$
-\frac{12}{23} $&$
0.4086 $&$ -0.4230 $&$ -0.8994 $&$ 0.1456 $\\
$h_i $&$ 2.2996 $&$ - 1.0880 $&$ - \frac{3}{7} $&$ -
\frac{1}{14} $&$ -0.6494 $&$ -0.0380 $&$ -0.0185 $&$ -0.0057 $\\
$e_i$ &$\frac{4661194}{816831}$&$ -\frac{8516}{2217}$ &$  0$ &$  0$ 
        & $ -1.9043$  & $  -0.1008$ & $ 0.1216$  &$ 0.0183$\\
$f_i$ & $-17.3023$ & $8.5027 $ & $ 4.5508$  & $ 0.7519$
        & $  2.0040 $ & $  0.7476$  &$ -0.5385$  & $ 0.0914$\\
$g_i$ & $14.8088$ &  $ -10.8090$  &$ -0.8740$  & $ 0.4218$ 
        & $  -2.9347$   & $ 0.3971$  & $ 0.1600$  & $ 0.0225$ \\
$l_i$ & $ 0.5784$ &  $-0.3921$  & $-0.1429$  & $ 0.0476$ 
        & $-0.1275$   & $0.0317$  & $0.0078$  & $-0.0031$ \\
\hline
\end{tabular}
\end{center}
\caption[]{Magic Numbers.
\label{tab:akh}}
\end{table}

\noindent
There are three scale uncertainties present in (\ref{main}):
\begin{itemize}
\item
The low energy scale $\mu_b=\ord(m_b)$ at which the Wilson
Coefficient $C_{7}^{(0){\rm eff}}(\mu_b)$ is evaluated.
\item
The high energy scale $\mu_W=\ord(\mw)$ at which 
the full theory is matched with the effective five-quark theory.
In LO this scale enters only $\eta$.
$C_{7}^{(0)}(\mu_W)$ and  $C_{8}^{(0)}(\mu_W)$, usually
denoted by
$C_{7}^{(0)}(\mw)$ and  $C_{8}^{(0)}(\mw)$, serve in LO as initial
conditions to the renormalization group evolution from $\mu_W$ down
to $\mu_b$. As seen explicitly in (\ref{c7}) and (\ref{c8}) they do
not depend on $\mu_W$.
\item
The scale $\mu_t=\ord(m_t)$ at which the running top quark mass is
defined. In LO it enters only $x_t$ in (\ref{xt}).
\end{itemize}
It should be stressed that $\mu_W$ and $\mu_t$ do not have to be
equal. Initially when the top quark and the W-boson are integrated
out, it is convenient in the process of matching to keep
$\mu_t=\mu_W$. Yet one has always the freedom to redefine the top
quark mass and to work with $\mtb(\mu_t)$ where $\mu_t\not=\mu_W$.
\noindent
It is evident from the formulae above that in LO the variations of
$\mu_b$, $\mu_W$ and $\mu_t$ remain uncompensated which results
in potential theoretical uncertainties in the predicted branching
ratio.

\noindent
In the context of phenomenological analyses of $B \to X_s\gamma$,
only the uncertainty due to $\mu_b$ has been discussed
\cite{AG1,BMMP:94,CZMM,GREUB}. It is the
purpose of this letter to analyze the uncertainties due to $\mu_W$
and $\mu_t$ and to reanalyze the $\mu_b$-uncertainty.

\noindent
It is customary to estimate the uncertainties due to $\mu_b$ by
varying it in the range $\mb/2\le\mu_b\le 2\mb$. Similarly one
can vary $\mu_W$ and $\mu_t$ in the ranges $\mw/2\le\mu_W\le 2\mw$
and $\mt/2\le\mu_t\le 2\mt$ respectively. Specifically in our
numerical analysis of $\mu_W$ and $\mu_t$ uncertainties 
we will consider the ranges
\be\label{ranges}
40~\gev\le\mu_W\le 160~\gev\qquad 80~\gev\le \mu_t\le 320\gev
\ee
setting $\mu_b=m_b\equiv m_{b,pole}=4.8\gev$.

\noindent
In the LO analysis we use
\begin{equation} 
\as(\mu) = \frac{\as(\mz)}{v(\mu)} 
\qquad v(\mu) = 1 - \beta_0 \frac{\as(M_Z)}{2 \pi} 
\ln \left( \frac{M_Z}{\mu} \right)
\label{eq:asmumz}
\end{equation} 
with $\alpha_s(\mz)=0.118$ and
\be\label{mbar}
\mtb(\mu_t)=\mtb(\mt)
\left[\f{\as(\mu_t)}{\as(\mt)}\right]^{\f{4}{\beta_0}}.
\ee
We are using the parameters $\alpha_s^{(5)}$ and
$\bar{m}^{(5)}$ defined in the effective theory  with five flavours 
throughout in this work, hence 
$\beta_0=23/3$. 
We set $\mtb(\mt)=168~\gev$
and $\mt\equiv m_{t,pole}=176~\gev$.

\noindent
Varying $\mu_W$ and $\mu_t$ in the ranges (\ref{ranges}) we
find the following  uncertainties in the branching
ratio:

\begin{equation}\label{LOmu1}
\Delta Br(B\to X_s \gamma)=\left\{ \begin{array}{ll}
\pm 13\% & (\mu_W) \\
\pm 3 \% & (\mu_t) \end{array} \right.
\end{equation}
to be compared with the $\pm 22\%$ uncertainty due to the variation
of the scale $\mu_b$ \cite{AG1,BMMP:94}. 
The fact that the $\mu_W$-uncertainty is smaller than
the $\mu_b$ uncertainty is entirely due to $\as(\mu_W)<\as(\mu_b)$. 
Still this uncertainty is rather disturbing as it introduces an error of
approximately $\pm 0.40\cdot 10^{-4}$ in the branching ratio.
The
smallness of the $\mu_t$-uncertainty is related to the weak $x_t$
dependence of $C_{7}^{(0)}(\mu_W)$ and  $C_{8}^{(0)}(\mu_W)$
which in the range of interest can be well approximated by
\be
C_{7}^{(0)}(\mu_W)=-0.122~ x_t^{0.30}
\qquad  C_{8}^{(0)}(\mu_W)=-0.072~ x_t^{0.19}
\ee
Thus even if $161\gev\le\mtb(\mu_t)\le 178\gev$ for $\mu_t$ in 
(\ref{ranges}),
the $\mu_t$ uncertainty in  $Br(B\to X_s \gamma)$ is small.
This should be contrasted with  $B_s\to\mu\bar\mu$,
$K_L\to\pi^0\nu\bar\nu$ and $ B^0-\bar B^0$ mixing, 
where $\mu_t$ uncertainties in LO have been
found \cite{BB2,BJW} to be $\pm 13\%$, $\pm 10\%$ and $\pm 9\%$ 
respectively.

\noindent
{\large\bf 4.}
We will next investigate how much the uncertainties in
(\ref{LOmu1}) are reduced after including NLO corrections.

\noindent
The formula (\ref{main}) modifies after the inclusion of NLO
corrections as follows \cite{CZMM}:
\be \label{ration}
\frac{Br(B \to X_s \gamma)}
     {Br(B \to X_c e \bar{\nu}_e)}
 = 
\frac{|V_{ts}^* V_{tb}|^2}{|V_{cb}|^2} 
\frac{6 \alpha}{\pi f(z)} F \left( |D|^2 + A \right)\,,
\ee
where
\be \label{factor}
F = \f{1}{\kappa(z)} 
\left( \f{\bar m_b(\mu=m_b)}{m_{b,{\rm pole}}} \right)^2 = 
    \f{1}{\kappa(z)} \left( 1 - \f{8}{3} \f{\as(m_b)}{\pi} \right),
\ee
with
$\kappa(z)$
being the QCD correction to the semileptonic decay
\cite{CM78} and given to a good approximation by \cite{KIMM}
\be \label{kap}
\kappa(z) = 1 - \frac{2 \as (\bar\mu_b)}{3 \pi}
\left[(\pi^2-\frac{31}{4})(1-\sqrt{z})^2+\frac{3}{2}\right] \,.
\ee
An exact analytic formula for $\kappa(z)$ can be found in \cite{N89}.
Here $\bar\mu_b=\ord(m_b)$ is a scale in the calculation of QCD corrections
to the semi-leptonic rate which is generally different from the one used
in the $b\to s\gamma$ transition. In this respect we differ from Greub et al.
\cite{GREUB}
who set $\bar\mu_b=\mu_b$. In \cite{CZMM} the choice
$\bar\mu_b=m_b$ has been made. We will return to this point below.

Next 
\be \label{Dvirt}
D = C_{7}^{(0){\rm eff}}(\mu_b) + \frac{\as(\mu_b)}{4 \pi} \left\{ 
C_{7}^{(1){\rm eff}}(\mu_b) + \sum_{i=1}^8 C_i^{(0){\rm eff}}(\mu_b) 
\left[ r_i + \frac{\gamma_{i7}^{(0){\rm eff}}}{2} \ln \frac{m^2_b}{\mu^2_b} 
\right] \right\}
\ee
where $C_{7}^{(1){\rm eff}}(\mu_b)$ is the NLO correction
to the effective Wilson coefficient of $Q_{7}$:
\be \label{C.expanded}
C^{\rm eff}_{7}(\mu_b) = C^{(0){\rm eff}}_{7}(\mu_b) + 
\frac{\as(\mu_b)}{4 \pi} C^{(1){\rm eff}}_{7}(\mu_b)\,. 
\ee
Generalizing the formula (21) of \cite{CZMM} to include $\mu_t$ and $\mu_W$
dependences we find\footnote{ We would like to thank the authors of
\cite{BorGre98} for pointing out the missing logarithmic term in the original 
version of this work. See also the discussion after equation (\ref{E3}).}
\bea     \label{c7eff1}
C^{(1)eff}_7(\mu_b) &=& 
\eta^{\f{39}{23}} C^{(1)eff}_7(\mu_W) + \f{8}{3} \left( \eta^{\f{37}{23}} 
- \eta^{\f{39}{23}} \right) C^{(1)eff}_8(\mu_W) 
\nonumber \\ &&
+\left( \f{297664}{14283} \eta^{\f{16}{23}}-\f{7164416}{357075} 
\eta^{\f{14}{23}} 
       +\f{256868}{14283} \eta^{\f{37}{23}} -\f{6698884}{357075} 
\eta^{\f{39}{23}} \right) C_8^{(0)}(\mu_W) 
\nonumber \\ &&
+\f{37208}{4761} \left( \eta^{\f{39}{23}} - 
\eta^{\f{16}{23}} \right) C_7^{(0)}(\mu_W) 
\nonumber \\ &&
+ \sum_{i=1}^8 \left(e_i \eta E(x_t) + f_i + g_i \eta
+\eta\left[
\frac{2}{3}e_i+6l_i
\right]\ln\frac{\mu_W^2}{M_W^2}
\right) \eta^{a_i},
\eea
where
in the $\overline{MS}$ scheme 
\be\label{GENC7}
C_7^{(1)eff}(\mu_W)= C_7^{(1)eff}(M_W)+
8 x_t \f{\partial C_7^{(0)}(\mu_W)}{\partial x_t}\ln\f{\mu_t^2}{\mw^2} 
+\left(\f{16}{3}C_7^{(0)}(\mu_W)-\f{16}{9} C_8^{(0)}(\mu_W)
+\f{\gamma_{27}^{(0){\rm eff}}}{2}\right) \ln \frac{\mu_W^2}{\mw^2} 
\ee
\be\label{GENC8}
C_8^{(1)eff}(\mu_W)= C_8^{(1)eff}(M_W)+
8 x_t \f{\partial C_8^{(0)}(\mu_W)}{\partial x_t}\ln\f{\mu_t^2}{\mw^2} 
+\left(\f{14}{3}C_8^{(0)}(\mu_W)
+\f{\gamma_{28}^{(0){\rm eff}}}{2}\right) \ln\frac{\mu_W^2}{\mw^2} 
\ee
Here ($x=x_t$)
\bea
C_7^{(1)eff}(M_W) &=& \f{-16 x^4 -122 x^3 + 80 x^2 -  8 x}{9 (x-1)^4} 
{\rm Li}_2 \left( 1 - \f{1}{x} \right)
                  +\f{6 x^4 + 46 x^3 - 28 x^2}{3 (x-1)^5} \ln^2 x 
\nonumber \\ &&
                  +\f{-102 x^5 - 588 x^4 - 2262 x^3 + 3244 x^2 - 1364 x +
208} {81 (x-1)^5} \ln x
\nonumber \\ &&
                  +\f{1646 x^4 + 12205 x^3 - 10740 x^2 + 2509 x - 436}
{486 (x-1)^4} 
\vspace{0.2cm} \\
C_8^{(1)eff}(M_W) &=& \f{-4 x^4 +40 x^3 + 41 x^2 + x}{6 (x-1)^4} 
{\rm Li}_2 \left( 1 - \f{1}{x} \right)
                  +\f{ -17 x^3 - 31 x^2}{2 (x-1)^5} \ln^2 x 
\nonumber \\ &&
                  +\f{ -210 x^5 + 1086 x^4 +4893 x^3 + 2857 x^2 - 1994 x
+280} {216 (x-1)^5} \ln x
\nonumber \\ &&
        +\f{737 x^4 -14102 x^3 - 28209 x^2 + 610 x - 508}{1296 (x-1)^4}
\eea
and
\be
E(x) = \frac{x (18 -11
x - x^2)}{12 (1-x)^3} + \frac{x^2 (15 - 16 x + 4 x^2)}{6 (1-x)^4} \ln
x-\frac{2}{3} \ln x.
\ee

The formulae for $C_{7,8}^{(1)eff}(M_W)$ given above and presented in 
\cite{CZMM} are obtained from
the results in \cite{Yao1,GH97} by using the general formulae for the
effective coefficient functions \cite{BMMP:94}. The formula for
$C_{7}^{(1)eff}(M_W)$
has been confirmed by us \cite{BKP2}.

\noindent
The numbers $e_i$--$l_i$ are given in Table \ref{tab:akh}. 
We have confirmed these numbers as well as the numerical coefficients
in (\ref{c7eff1}) using the anomalous dimension matrices  in
\cite{CZMM}. 
Next the $\eta$  in
(\ref{ET}) should now be calculated using the NLO expression
\be \label{alphaNLL}
\as(\mu) = \frac{\as(M_Z)}{v(\mu)} \left[1 - \f{\beta_1}{\beta_0} 
           \frac{\as(M_Z)}{4 \pi}    \f{\ln v(\mu)}{v(\mu)} \right],
\ee
where 
$v(\mu)$ is given in (\ref{eq:asmumz}) and
$\beta_1 = \frac{116}{3}$.

The constants $r_i$ resulting from the calculations of NLO corrections
to decay matrix elements \cite{GREUB} are collected in \cite{CZMM}
where also explicit formulae for $C_i^{(0){\rm eff}}(\mu_b)$
with $i=1-6,8$ and the values of $\gamma_{i7}^{(0){\rm eff}}$ can be found.
It should be stressed that the basis of the operators with $i=1-6$ used
in \cite{CZMM} differs from the standard basis used in the literature
\cite{BBL}.
For the discussion below it will be useful to have \cite{CFMRS:93}
\be
\gamma_{27}^{(0){\rm eff}}=\f{416}{81} \qquad
\gamma_{28}^{(0){\rm eff}}=\f{70}{27}
\ee
which enter (\ref{GENC7}) and (\ref{GENC8}) respectively.

\noindent
Finally the term $A$ in (\ref{ration}) originates from the bremsstrahlung
corrections and the necessary virtual corrections needed for the
cancellation of the infrared divergences. These have been
calculated in \cite{AG2,Pott} and are also considered in 
\cite{CZMM,GREUB} in the
context of the full analysis. The explicit formula for A, which we use
in our numerical analysis, can be found in equation (32) of 
\cite{CZMM}\footnote{In the replacement version of \cite{CZMM}
several quantities entering the formula for A have been corrected.
In this paper the updated values are used. 
We thank M.~Neubert \cite{Neu98}, P.~Gambino and M.~Misiak for
informing us about these modifications.
Accordingly the numerical results
of this work are changed slightly.}.

\noindent
Setting $\mu_W=\mu_t=\mu_{Wt}$, replacing $\gamma_{27}^{(0){\rm eff}}$
by its value in the NDR scheme 
$\gamma_{27}^{(0){\rm NDR}}=464/81$ and adding all $\mu_i$ dependent
terms in (\ref{GENC7}) we recover the $\mu_{Wt}$ dependence of
$C_7^{(1){\rm NDR}}(\mu_{Wt})$ found in \cite{GH97}. Similarly 
replacing $\gamma_{28}^{(0){\rm eff}}$
by $\gamma_{28}^{(0){\rm NDR}}=76/27$ in (\ref{GENC8}) we recover
the $\mu_{Wt}$ dependence of
$C_8^{(1){\rm NDR}}(\mu_{Wt})$ given in \cite{GH97}.
For $\mu_W=\mu_t=\mw$ the formulae above reduce to the ones given
in \cite{CZMM}.

\noindent
{\large \bf 5.}
Before entering the numerical analysis let us demonstrate analytically
that the $\mu_t$ and $\mu_W$ dependences present in 
$C^{(0){\rm eff}}_{7}(\mu_b)$ are indeed cancelled at $\ord(\as)$
by the explicit scale dependent terms in (\ref{GENC7}). The scale
dependent terms in (\ref{GENC8}) do not enter this cancellation
at this order in $\as$ in $B\to X_s \gamma$. On the other hand
they are responsible for the cancellation of the scale dependences
in $C^{(0){\rm eff}}_{8}(\mu_b)$ relevant for the $b \to s~{\rm gluon}$
transition. 

\noindent
Expanding the three terms in (\ref{C7eff}) in $\as$ and
keeping the leading logarithms we find:
\be\label{E1}
\eta^\frac{16}{23} C_{7}^{(0)}(\mu_W)=
\left (1+\f{\as}{4\pi}\f{16}{3}\ln\f{\mu_b^2}{\mu^2_W}\right)
C_{7}^{(0)}(\mu_W)
\ee

\be\label{E2}
 \frac{8}{3}
   \left(\eta^\frac{14}{23} - \eta^\frac{16}{23}\right) 
C_{8}^{(0)}(\mu_W)= 
-\f{\as}{4\pi}\f{16}{9}\ln\f{\mu_b^2}{\mu^2_W}
C_{8}^{(0)}(\mu_W)
\ee

\be\label{E3}
\sum_{i=1}^8 h_i \eta^{a_i}= \f{\as}{4\pi}\f{23}{3}\ln\f{\mu_b^2}{\mu^2_W}
\sum_{i=1}^8 h_i a_i =\f{208}{81} \f{\as}{4\pi}\ln\f{\mu_b^2}{\mu^2_W}
\ee
respectively. In (\ref{E3}) we have used $\sum h_i=0$. 
Inserting these expansions into (\ref{Dvirt}),
we observe that the $\mu_W$ dependences
in (\ref{E1}), (\ref{E2}) and (\ref{E3})
are precisely cancelled  by the three  explicit logarithms in
(\ref{GENC7}) involving $\mu_W$, respectively. Similarly one can convince
oneself that the $\mu_t$-dependence of $C^{(0){\rm eff}}_{7}(\mu_b)$
is cancelled at $\ord(\as)$ by 
the $\ln \mu_t^2/\mw^2$ term in (\ref{GENC7}).

Interestingly
the last logarithm in (\ref{c7eff1}) does not contribute to
any cancellation of the $\mu_W$ dependence
at this order in $\as$ due to the relation
$
\sum_{i=1}^8 \left(\frac{2}{3}e_i  + 6 l_i \right)=0
$
which can be verified by using the Table \ref{tab:akh}.

\noindent
Clearly there remain small $\mu_t$ and $\mu_W$ dependences in
(\ref{ration}) which can only be reduced by going beyond the NLO
approximation. They constitute the theoretical uncertainty which
should be taken into account in estimating the error in the
prediction for $Br(B\to X_s\gamma)$.

\noindent
Using the well known two-loop generalization of $(\ref{mbar})$
and varying $\mu_W$ and $\mu_t$ in the ranges (\ref{ranges}) we
find that the respective uncertainties in the branching
ratio after the inclusion of NLO corrections are negligible:
\begin{equation}\label{LOm}
\Delta Br(B\to X_s \gamma)=\left\{ \begin{array}{ll}
\pm 1.1\% & (\mu_W) \\
\pm 0.4\% & (\mu_t) \end{array} \right.
\end{equation}

\noindent
{\large\bf 6.} 
We have next performed the NLO analysis of the $\mu_b$ dependence.
Varying $\mu_b$ in the range $2.5\gev\le\mu_b\le 10\gev$ we find
\begin{equation}\label{NLOm}
\Delta Br(B\to X_s \gamma)=\pm 4.3\%\quad (\mu_b)
\end{equation}
This reduction of the $\mu_b$-uncertainty by roughly a factor of five 
 relative to $\pm 22\%$ in LO is caused by the presence of
the explicit logarithm  $\ln{m^2_b}/{\mu^2_b}$ in (\ref{Dvirt}). 
We note that our result in (\ref{NLOm}) differs from
 the $\mu_b$- uncertainty of $\pm 6.6\%$  quoted in
\cite{CZMM}. A discussion with the latter authors confirmed our result.

Next we would like to comment on the uncertainty due to variation of
$\bar\mu_b$ in $\kappa(z)$ given in (\ref{kap}). In \cite{GREUB}
the choice $\bar\mu_b=\mu_b$ has been made. Yet in our opinion
such a treatment is not really correct, since the scale $\bar\mu_b$ in
the semi-leptonic decay has nothing to do with the scale $\mu_b$
in the renormalization group evolution in the $B\to X_s\gamma$
decay. 
Varying $\bar\mu_b$ in the range $2.5\gev\le\mu_b\le 10\gev$ we find
\begin{equation}\label{NLOm1}
\Delta Br(B\to X_s \gamma)=\pm 1.7\% \quad (\bar\mu_b)
\end{equation}
Since the $\mu_b$ and $\bar\mu_b$ uncertainties are uncorrelated we
can add them in quadrature finding $\pm 4.6\%$ for the total
scale uncertainty due to $\mu_b$ and $\bar\mu_b$. 
This is smaller by roughly 30\% than the case in which $\bar\mu_b=\mu_b$
is used.
The addition of the uncertainties in $\mu_t$ and $\mu_W$ in
(\ref{LOm}) modifies this result slightly and the total scale 
uncertainty in $Br(B \to X_s\gamma)$ amounts then to
\be\label{stheon}
\Delta Br(B{\to}X_s \gamma) = \pm 4.8\% \quad({\rm scale})
 \ee
which is roughly by a factor of 1.5 smaller than quoted
in \cite{CZMM,GREUB}.

It should be stressed that this pure theoretical 
uncertainty related to the truncation of the perturbative series
should be distinguished from parametric uncertainties related
to $\alpha_s$, the quark masses etc. discussed below.

In our numerical calculations we have included all corrections
in the NLO approximation. To work  consistently
in this order, we have in particular
expanded the various factors in (\ref{ration}) in  $\alpha_s$ and discarded
all NNLO terms of order $\alpha_s^2$ which resulted in the process
of multiplication. This treatment is different
from \cite{CZMM,GREUB}, where the $\alpha_s$ corrections in (\ref{factor}) 
have not been expanded in the evaluation of 
(\ref{ration}) and therefore some higher order corrections have been kept.
Different scenarios of partly incorporating higher order corrections
by expanding or not expanding various factors in (\ref{ration})
affect the branching ratio by $\Delta Br(B\rightarrow X_s\gamma)\approx
\pm 3.0 \%$. This number indicates that indeed the scale uncertainty
in (\ref{stheon}) realistically  estimates the magnitude of yet
unknown higher order corrections.   

The remaining uncertainties are due to the values of the various 
input parameters.
In order to obtain the final result for the branching ratio
we have used  the same parameters as in \cite{CZMM}.
They are given in Table~2.
\begin{table}[htb]
\begin{center}
\begin{tabular}{|c|c|c|c|c|c|c|c|}
\hline
 & $\as(M_Z)$ & $m_{t,pole}$   & $m_{c,pole}/m_{b,pole}$ 
& $m_{b,pole}$ & $\alpha_{em}^{-1}$ 
& $|V_{ts}^{\star}V_{tb}|/V_{cb}$ & $Br(B\to X_c e\bar\nu_e)$ \\
\hline
{\rm Central} & $0.118$ & $176$  & $0.29$ & $4.8$  &  
  $130.3$  & $0.976$ &$0.104$ \\
\hline
{\rm Error} & $\pm 0.003$   & $\pm 6.0$  & $\pm 0.02$ & $\pm 0.15$  &  
  $\pm 2.3$  & $\pm 0.010$ &$\pm 0.004$ \\
\hline
\end{tabular} 
\vspace{0.2cm}\\
{\small Table 2. Input parameter values and their uncertainties.
The masses are given in GeV.
}
\end{center}
\end{table}
In addition  we have included small $1/m_b^2$ corrections
as in \cite{CZMM} and also a $3\%$ enhancement \cite{BUC97}
 from $1/m_c^2$ corrections
\cite{VOL96}\footnote{In contrast to a 3\% suppression found
originally in
\cite{VOL96} (except for the second paper which actually discusses
the exclusive channels),
 the $1/m_c^2$ corrections to the $B\rightarrow
X_s\gamma$ decay have been shown to be positive in \cite{BUC97}.}
 which were not known at the time of the analysis
\cite{CZMM}.
The relative importance of various
uncertainties is shown in Table~3. Comparing this table with
the first row in the corresponding table in \cite{CZMM} we
observe that except for the scale uncertainties discussed
above, our error analysis agrees well with the one presented
in \cite{CZMM}.
\begin{table}[htb]
\begin{center}
\begin{tabular}{|c|c|c|c|c|c|c|c|}
\hline
{\rm Scales} & $\as(M_Z)$ & $m_{t,pole}$  & $m_{c,pole}/m_{b,pole}$ 
& $m_{b,pole}$ & $\alpha_{em}$ & CKM angles & $B\to X_c e\bar\nu_e$ \\
\hline
$\pm 4.8\%$ & $\pm 2.9\%$   & $\pm 1.7\%$  & $\pm 5.4\%$ & $\pm 0.7\%$  &  
  $\pm 1.8\%$  & $\pm 2.0\%$ &$\pm  3.8\%$ \\
\hline
\end{tabular} 
\vspace{0.2cm}\\
{\small Table 3. Uncertainties in $Br(B \to X_s\gamma)$ due to various 
sources.}
\end{center}
\end{table}

Adding all the uncertainties 
in quadrature we find  
\be\label{sfin}
Br(B{\to}X_s \gamma) =(3.60 \pm 0.17~({\rm scale})~\pm 0.28~({\rm par})) 
  \times 10^{-4}
= (3.60 \pm 0.33)  \times 10^{-4}
\ee
where we show separately scale and parametric uncertainties.
 
Comparing this result with the one of \cite{CZMM} as given
in  (\ref{theon}) we observe that in spite of the smaller
scale uncertainties in (\ref{sfin}) our final result is compatible
with the one of \cite{CZMM} and the one given in \cite{GREUB}. This is due
to the parametric uncertainties which dominate the theoretical
error at present. Once these parametric uncertainties will be reduced
in the future the smallness of the scale uncertainties achieved
through very involved QCD calculations, in particular in 
\cite{CZMM,GREUB,AG2,Pott,Yao1,GH97,BKP2}, can be better appreciated.
This reduction of the theoretical error in the Standard Model
prediction for $Br(B{\to}X_s \gamma)$ could turn out to be very
important in the searches for new physics. To this end also a better
understanding of non-perturbative corrections \cite{LDGAMMA} beyond 
those considered
here should be achieved.

The theoretical estimate in (\ref{sfin}) is somewhat higher than
the CLEO result in (\ref{EXP}) and rather close to the ALEPH result
in (\ref{EXP2}). In any case we conclude that within
 the remaining theoretical
and in particular experimental uncertainties, the Standard Model value
is compatible with experiment. It will be exciting to watch the
improvements in the theoretical estimate and in the experimental value
in the coming years.

{\bf Acknowledgments}

We would like to thank Mikolaj Misiak and Manfred M{\"u}nz for
a useful discussion concerning their numerical analysis.

\vfill\eject

\end{document}